\documentclass[twocolumn,showpacs,preprintnumbers,amsmath,amssymb]{revtex4}
\usepackage{graphicx}
\usepackage{dcolumn}
\usepackage{bm}

\newcommand{\be}{\begin{equation}}
\newcommand{\ee}{\end{equation}}
\newcommand{\ba}{\begin{eqnarray}}
\newcommand{\ea}{\end{eqnarray}}

\begin{document}\title{On the  quark-gluon plasma search}
\author{S.D. Hamieh}
\affiliation{\it Kernfysisch Versneller Instituut, Zernikelaan 25, 9747 AA Groningen, The Netherlands.}
\date{\today}

\begin{abstract}
We report on the effect of the quantum statistics on the two-proton spin correlation (SC) in cold and thermal nuclear matter. We have found that two nucleons SC function can be well approximated by a guassian  with correlations length $\sigma\sim1.2$ fm. 
We have proposed SC measurement on low protons energy  as test of the quark-gluon plasma formation in relativistic heavy
ions collisions. 
  
\end{abstract}
\maketitle

Lattice QCD predict a phase transition from Hadrons gaz (HG) to quark-gluon plasma (QGP) at deconfinement temperature, $T_d\sim 170\,$MeV. It is believed that  $T_d$ has been reached in the relativistic heavy ions collision. 
Several signatures of the QGP formation have been proposed in the literature. The proposed signatures include:
\begin{itemize}
\item Dilepton production;
\item Photon production;
\item Hanbury-Brown-Twiss Effect;
\item Strangeness enhancement;
\item $J/\Psi$ suppression. 
\end{itemize} 
All this effect have been intensively studied  in the literature. 
For example, strangeness enhancement \cite{raf1} has been observed at the SPS energy and recently at RHIC. However, it has been shown that strangeness enhancement can be explained in terms of statistical model formulated in
canonical ensemble with respect to strangeness conservation \cite{ham2}. This problem also hold for other signatures in the sense that
those signatures can be considered as clear indication in the favor of QGP formation but not decisive.
Therefore, one may ask the following question: is there exists a signature that shows up {\it only} in one phase  (QGP or HG) 
and not in the other  one?  This will be our goal.
In this paper, we propose the spin correlation (SC) 
measurement as fundamental probe. SC discussed here is
due to the indistinguishability of nucleons and the existence of quantum statistic discussed, in different context,
 in \cite{6}.

It is known that if the QGP is produced in relativistic heavy ions collision a supercooling and sudden hadronization is
expected \cite{ham3}. Therefore, in this scenario, the hadrons are produced from phase space. Thus, from this point of view,
SC due to the quantum statistics will not play a role as the hadrons produced suddenly and escape from the local volume. However, if the hadrons
are produced from HG phase they have to respect quantum statistics. Therefore, a possible SC can be expected. To the best of our knowledge, for a non-comprehensible reason, this effect has
 been rarely studied explicitly in nuclear physics \cite{bert1}.

In the next section we aim on the evaluation of the SC function, the quantum,
and the classical correlations between two protons inside a nucleus. 
In section II we will discuss the two-proton SC function 
in thermal nuclear matter as test of QGP formation. Finally our conclusion
is given in section III.

\section{\it SC in cold nuclear matter} 
\begin{figure}\includegraphics{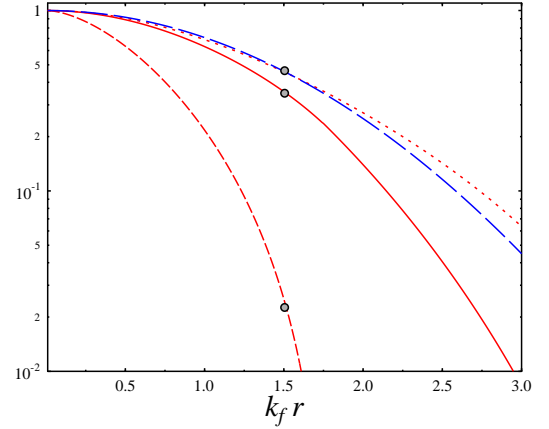}
\vspace{-0.7cm}\caption{The correlation function at $0^{\circ}$ (dotted line), the relative entropy of entanglement  (dashed line) and the classical correlations (solid line) are depicted in dependence on $k_fr$. Small circles  represent our predictions for the nuclear matter where we have used  $k_f\sim 1.27 \,{\rm fm}^{-1}$ and $r\sim 1.2$ fm.\protect\label{werner}. The large dashed line represent the gaussian approximation of the mixing parameter.}
\end{figure}

In this section, we will evaluate the correlations between two nucleons (protons or neutrons) in Fermi sea approach to nuclear matter \cite{101,100}.
 Therefore, in the coordinate representation, the density matrix elements of two nucleons are

\be \langle\Psi_0|\psi^{\dag}_{\sigma_2^{\prime} \bf r_2^{\prime}}\psi^{\dag}_{\sigma_1^{\prime} \bf r_1^{\prime}}\psi_{\sigma_2 \bf r_2}\psi_{\sigma_1 \bf r_1}|\Psi_0\rangle\,,\ee
where $\psi_{\sigma \bf r}=\frac{1}{V}\sum_{\bf k}e^{i\bf k\cdot r}a_{\sigma \bf k}$ and 
$|\Psi_0\rangle=\prod_{|k|\leq k_f}a^{\dag}_{\sigma \bf k}|0\rangle\,$ is the  ground state of the system. 
As usual, in the continuum limits $[a^{\dag}_{\sigma \bf k},a_{\sigma^{\prime} \bf k^{\prime}}]=\delta_{\sigma\sigma^{\prime}}\delta({\bf k- k^{\prime}})$.

A straightforward calculations gives the explicit form of the two-particle space-spin density matrix \cite{99}
\ba \rho^{sp}&=&\frac{1}{2}\left(F({\bf r_1- r_1^\prime})F({\bf r_2 - r_2^\prime})\delta_{\sigma_1\sigma_1^{\prime}}\delta_{\sigma_2\sigma_2^{\prime}}
\nonumber\right.\\&&\left.-F({\bf r_1- r_2^\prime})F({\bf r_2 - r_1^\prime})\delta_{\sigma_1\sigma_2^{\prime}}\delta_{\sigma_2\sigma_1^{\prime}}
 \right)\,\ea
with $F({\bf r})=\frac{1}{V}\sum_{\bf k}e^{i\bf k\cdot r}$. Depending on the space density matrix, two spins may be entangled. We obtain Vedral's
result \cite{6} only if ${\bf r_1 = r_1^\prime}$ and  ${\bf r_2 = r_2^\prime}$, that is, only diagonal elements of a space density matrix are considered.
The two-spin density matrix, depending on the relative distance between two nucleons $r = |{\bf r_1 - r_2}|$, reads
\ba\rho^{s}=\frac{1}{4-2f^2}\left(\begin{array}{cccc}1-f^2&0&0&0\\
0&1&-f^2&0\\
0&-f^2&1&0\\
0&0&0&1-f^2\end{array}\right)\,,
\ea
 $f(r)= 3j_1(k_f r)/k_f r$ and $j_1$ is the spherical Bessel function. 
 We find that $\rho^{s}$ is a Werner state characterized by a single parameter $p=f^2/(2-f^2)$.

\begin{figure}\includegraphics{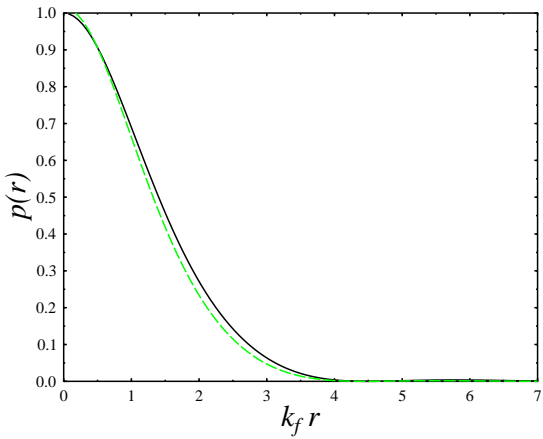}
\vspace{-0.7cm}\caption{The mixing parameter $p(r)$  at $T=0$ (solid line)
and for $T=0.2\mu_B$ (dashed line).\label{fig1}}
\end{figure}

In  Fig.~\ref{werner} we show as function of $k_f r$ the correlation function at $0^{\circ}$ \cite{ham4}  
\be P(\theta)={N_{++}+N_{--}-N_{+-}-N_{-+} \over N_{total}}=p\cos(\theta)\,,\ee 
($\theta$ is the angle between the quantization axis and $+$ and $-$ represent the positive and negative spin projection on the quantization axis) 
and  the relative entropy of entanglement~\cite{Vedr97}, 
\be E_r=\min_{\rho^{*}\in {\cal D}}S(\rho\|\rho^{*})\label{eq:ccorr}\,,\ee
where ${\cal D}$ is a set of all separable states in the Hilbert space in which $\rho$ is defined. Also
shown in  Fig.~\ref{werner} the classical correlations~\cite{Hami03}, 
\be \Psi(\rho)=S(\rho\|\rho_A\otimes\rho_B)-  \min_{\rho^{*}\in {\cal D}}S(\rho\|\rho^{*})\,, \label{eq:qcorr} \ee
where $\rho_A$ and  $\rho_B$ are the  reduced density matrices. The small circles in  Fig.~\ref{werner} represent our predictions for the nuclear matter with typical values of $k_f\sim 1.27 \,{\rm fm}^{-1}$ and $r\sim 1.2$ fm \cite{100,101}. We
observe that the  mixing parameter, $p(r)= P(0^{\circ})$, 
can be approximated with a gaussian function 
\be p(r)=e^{-r^2/2\sigma^2}\,,\ee
with $\sigma=1.2$ fm, see Fig. 1.

The mixing parameter, $p(r)$, can be measured experimentally. In fact, following Bertulani  approach \cite{bert1} for weakly bound nucleons as
in $\rm ^1H(^6He,^2He)^5H$ reaction one can argue, using our results, that the relative energy distribution of the two outgoing proton $\rm ^2He$
without the final state interaction should reads

\be\sigma(E_{pp})\propto \sqrt{E_{pp}}\left(1+e^{-R^2/2\sigma^2}[\cos(\sqrt{m_nE_{pp}/\hbar^2}R)]\right)
\,,
\ee
where $E_{pp}$ is the relative energy between the two protons and $R$ is the size  of the source region. Therefore, from experimental data \cite{kors1}
we can extract the mixing parameter $p(r)$ \cite{hami3}. Also our mixing parameter can be extracted from coulomb dissociation of $\rm ^{11}Li$ 
\cite{simo1} see forthcoming paper \cite{hami3}.

\section{\it SC in thermal nuclear matter} 
For simplicity we will only consider  protons from the all known particles and resonances of HG. The formulation described in the above section can be applied in this case. However, at finite temperature the occupation probability of state ${\bf k}$ is the Fermi Dirac distribution 
$n_{\bf k}=(1+e^{\beta(\epsilon_{\bf k}-\mu_B)})^{-1}$. As results the function $f(r)$ should be replaced by:
\be f(r)=2\frac{\sum_{\bf k}n_{\bf k}e^{i\bf k \cdot r}}{\sum_{\bf k}n_{\bf k}}.\ee

From the last section we know that $P(0^{\circ})=p=f^2/(2-f^2)$. Thus, in order to find the correlations function, we need an estimation of $f$.
Because, the thermal freeze out temperature is $T_{th}\sim 100$MeV well below the baryonic chemical potential $\mu_B\sim 266$ MeV \cite{stac1}, the correlation function can be approximated by $p(r,T=0)$ as it can be seen in Fig. \ref{fig1}. 
However, with this conditions $k_f\sim 10\,\rm fm^{-1}$, thus the correlation function is zero. Nevertheless, we can still 
have correlation if we consider only the outgoing proton with momentum {$\rm <2\,fm^{-1}$} which produce a large
amount of correlation $\sim 0.25$ that can be measured experimentally. 
 For example, in 158 AGeV Pb--Pb collisions at SPS,  assuming Bjorken scaling
 at the central rapidity with $\Delta y\sim 0.2$, the longitudinal dimension of the volume is $d=\Delta y \,\tau \sim 2\,$fm. Thus, SC can be performed on those protons. A carbon analyzer as in \cite{ham4} is suitable for this analysis. 

Moreover, our results can have effect on the two proton Hanbury-Brown-Twiss (HBT) interferometery.  In fact, it is known that the two-proton correlation function is influenced by identical particle interference, short-range hadronic interaction, and long-range Coulomb repulsion.
For noninteracting identical particles, the squared wave function has the form

\be \Psi^2({\bf q,r})\propto 1\pm \cos({2\bf qr})\,,\ee
where the plus sign stands for singlet spin state and minus sign for triplet. In the literature two assumption are
frequently used \cite{wolf1}
\begin{itemize}
\item spin parallel orientations (triplet) or,
\item random spin orientations.
 \end{itemize}
Clearly from the discussion above the situation can be different and contribution from singlet state can dominate
for small frequency \cite{hami3}. This effect can be seen in HBT interferometery by considering large relative energy in order to reduce the final state interaction
\cite{na491,hami3}.

\section{Conclusion}
In this paper, we have proposed two-proton spin correlation measurement as potential test of the quark-gluon
plasma formation in relativistic heavy ions collisions. We have found  that a non negligible correlation can be measured
experimentally for protons with momentum {\rm $\rm <2\,fm^{-1}$}.
 A more detailed calculations will be addressed in the near future.
\section*{Acknowledgments}
This work was performed as part of the research program of the {\sl Stichting voor Fundamenteel Onderzoek der Materie (FOM)}
with financial support from the {\sl Nederlandse Organisatie voor Wetenschappelijk Onderzoek }.



\begin{thebibliography}{99}
%



\bibitem{raf1}J. Rafelski, and B. Muller,  {\it Phys Rev Lett} {\bf 48}, 1066, (1982)
\bibitem{ham2}S. Hamieh, K. Redlich, and A. Tounsi, {\it Phys.Lett. B} {\bf 486}, 61, (2000)
\bibitem{6}V. Vedral, {\it Central Eur.J.Phys.}{\bf 1}, 289 (2003).
\bibitem{bert1} C. Bertulani, {\it J. Phys. G} {\bf 29}, 769, (2003)


\bibitem{ham3} S. Hamieh, J. Letessier, and J. Rafelski, {\it Phys.Rev. C} {\bf 62}, 064901, (2000)
\bibitem{99}C. N. Yang, {\it Rev. Mod. Phys. }{\bf 34}, 694 (1962)

\bibitem{Vedr97} V. Vedral, {\it et al.},{\it
 Phys. Rev. Lett.}  {\bf 78}, 2275 (1997)

\bibitem{Hami03} S. Hamieh, {\it et al.}, {\it
    Phys. Rev. A} {\bf 67}, 014301 (2003)
\bibitem{100}
Shell model and optical potential suggest that, to a first approximation, we may picture the nucleus as potential well with 42 MeV deep and filed up to about 8 MeV below the top. Thus, $k_f\sim 1.27\,{\rm fm}^{-1}$ and $r\sim 1.2$ fm.
\bibitem{101}
Similar calculations can be made in Shell model.



\bibitem{kors1}A. Korsheninnikov {\it et. al}, {\it  Phys. Rev. Lett.} {\bf 87}, 092501 (2002)


\bibitem{hami3}S. Hamieh,  {\it in preparation}

\bibitem{simo1} H. Simon,  {\it et al.}, {\it  Phys. Rev. Lett.} {\bf 83}, 496 (1999)
\bibitem{stac1} P. Braun-Munzinger, I. Heppe, and J. Stachel, {\it 
Phys.Lett. B} {\bf 465}, 15, (1999)

\bibitem{ham4}S. Hamieh, {\it et al.}, {\it
J. Phys. G} {\bf 30}, 481, (2004)
\bibitem{wolf1} W. Bauer, and C. Gelbke,  {\it Annu. Rev. Nucl. Part. Sci. } {\bf 42}, 77, (1992)

\bibitem{na491}NA49 Collaboration, {\it Phys. Lett. B} {\bf 467}, 21, (1999)
\end{thebibliography}
\end{document}